\documentclass[twocolumn,letterpaper,amsmath,amssymb,floatfix,aps,superscriptaddress]{revtex4}
\usepackage{graphicx}
\usepackage{dcolumn}
\usepackage{bm}
\usepackage{epsfig,color}

\newcommand{\Av}[1]{{\bf #1}}

\def\ln{{\operatorname{ln}}}

\def\rmd{{\mathrm{d}}}

\def\rme{{\mathrm{e}}}
\def\lB{\ell_{\mathrm{B}}}

\begin{document}

\title{Dressed Counterions: Poly- and Monovalent Ions at Charged Dielectric Interfaces}

\author{Matej Kandu\v c}
\affiliation{Department of Physics, Technical University of Munich, 85748 Garching, Germany}
\affiliation{Department of Theoretical Physics, J. Stefan Institute, SI-1000 Ljubljana, Slovenia}

\author{Ali Naji}
\affiliation{Department of Applied Mathematics and Theoretical Physics, University of Cambridge, Cambridge CB3 0WA, United Kingdom}

\author{Jan Forsman}
\affiliation{Department of Theoretical Chemistry, Lund University Chemical Center, P.O. Box 124, S-221 00 Lund, Sweden}


\author{Rudolf Podgornik}
\affiliation{Department of Theoretical Physics,
J. Stefan Institute, SI-1000 Ljubljana, Slovenia}
\affiliation{Department of Physics, Faculty of Mathematics and Physics and Institute of Biophysics, Medical Faculty,
University of Ljubljana, SI-1000 Ljubljana, Slovenia}

\begin{abstract}
We investigate the ion distribution and overcharging at charged interfaces with dielectric inhomogeneities in the presence of asymmetric electrolytes containing  polyvalent and monovalent ions. We formulate an effective  ``dressed counterion" approach by integrating out the monovalent salt degrees of freedom and show that it agrees with results of explicit Monte-Carlo simulations. We then  apply the dressed counterion approach within the framework of the strong-coupling theory, valid for polyvalent ions at low concentrations, which  enables  an  analytical description for salt effects as well as dielectric inhomogeneities in the limit of strong Coulomb interactions on a systematic level.  Limitations and applicability of this theory are examined by comparing the results with simulations. 
\end{abstract}
\maketitle


\thispagestyle{plain}
\pagestyle{plain}

\pagenumbering{arabic}
\setlength\arraycolsep{2pt}

\section{Introduction}
Coulomb interactions are important in almost all soft matter systems at room temperature~\cite{leshouches}. Salient features of Coulomb systems such as ionic solutions arise as a consequence of 
the extreme long range of these interactions. In the last decade, concerted efforts were invested in order to understand 
the statistical behavior of soft matter systems in the presence of polyvalent ions. It became evident that such ions can behave very differently from monovalent ones 
\cite{newrev,Kjellander,shklovskii,Levin,Netz,hoda,naji}. The reason lies in different strengths of electrostatic interactions between constituents of multi-component ionic solutions. The distance at which the interaction energy between $q$-valent ions equals the thermal energy $k_\textrm{B}T$ is $q^2\lB$, where $\lB=e_0^2/(4\pi\varepsilon\varepsilon_0 k_\textrm{B}T)$ is known as the Bjerrum length (with $e_0$ being the elementary charge and  and  $\varepsilon$ the dielectric constant of the solvent). Similarly, the distance at which each ion interacts with an oppositely charged surface  (of absolute surface charge density $\sigma$) with an energy scale  $k_\textrm{B}T$ is the Gouy-Chapman length, $\mu=e_0/(2\pi q\lB\sigma)$. The behavior of a charged system depends on the ratio of the two, which defines an {\em electrostatic coupling parameter} $\Xi=q^2 \lB/\mu$~\cite{Netz}. In many practical cases for monovalent ions this coupling parameter is $\Xi\lesssim 1$, which justifies a mean-field Poisson-Boltzmann (PB) description of the system. In contrast, for polyvalent ions $\Xi\gg 1$, which leads to diverse phenomena, such as overcharging and like-charge attraction~\cite{Nguyen1, Nguyen2,Nguyen3,newrev,Kjellander,shklovskii,Levin,Netz,hoda,naji}. 

A strong-coupling (SC) theory valid for $\Xi\gg 1$ provides a consistent and systematic theoretical framework for understanding Coulomb systems in this limit. Until now it has been formulated explicitly for a counterion-only system~\cite{hoda}, which has been extended to a case with dielectric inhomogeneities as well~\cite{jho-prl}. We now make a crucial step forward (in order to study the more realistic and experimentally relevant cases) and formulate the SC theory for a mixture of symmetric and asymmetric  electrolytes (poly- as well as monovalent salts) in the presence of dielectric inhomogeneities. 

In order to investigate the salient features and to test our theoretical approach to highly asymmetric multicomponent Coulomb systems we set ourselves to study the polyion concentration profiles and the possibility of overcharging by three widely different approaches. The first one is the {\em standard (explicit) Monte Carlo (MC) simulation}, which treats all ions explicitly, and will merely serve as a test for the other two approaches. In the second, {\em dressed counterion approach}, we integrate out the monovalent ion degrees of freedom by replacing the Coulomb interaction between remaining charges with an effective screened Debye-H\"uckel (DH) interaction (as first described in Ref.~\cite{kanduc-dressed} for a dielectrically  homogeneous system with uniform screening). Such screened interactions are then used to perform {\em implicit MC simulations} that treat all charges explicitly {\em except} the monovalent ions, and show an extremely good agreement with explicit MC, confirming the validity of the approach based on dressed counterions in highly asymmetric systems $q\gg 1$. Finally, we implement the {\em dressed counterions strong coupling theory}~\cite{kanduc-dressed}, which  allows us to obtain an analytical description of ionic distributions and overcharging in the limit of strong coupling with the local surface field.
 The validity of the SC dressed counterion theory is thus examined by making a comparison with both explicit and implicit simulations. 

\section{The model}

\begin{figure}[t]
\centerline{\psfig{figure=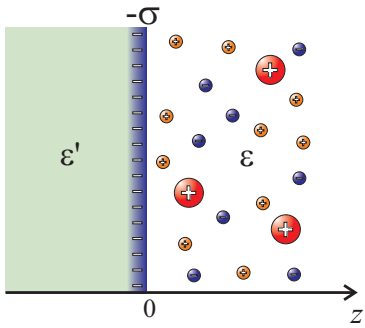,width=4.8cm}}
\caption{\label{fig:geom}
(Color online) Schematic view of an impenetrable charged dielectric interface of surface charge density $-\sigma$ bathed by a monovalent 1:1 salt and polyvalent $q$:1 salt, with larger $q$-valent counterions. For water-hydrocarbon interface one has $\varepsilon = 80$, $\varepsilon' = 2$, which gives dielectric $\Delta=0.95$ that allows for an approximate treating as $\Delta=1$.}
\end{figure}

The model system considered here, Fig.~\ref{fig:geom}, comprises an impenetrable  charged planar dielectric interface with surface charge density $-\sigma$ bathed in a mixture of monovalent 1:1 and polyvalent $q$:1 salts with bulk concentrations of $n_0$ and $c_0$, respectively. The total bulk concentration of monovalent ($+1$) counterions is therefore $n_0$, that of polyvalent ($+q$) counterion concentration $c_0$, and monovalent ($-1$) coion concentration $n_0+q c_0$. We shall refer to $+q$ polyvalent counterions simply as ``counterions'', which should not be confused with monovalent salt counterions. We will express the bulk concentration of polyvalent counterions by the parameter 
\begin{equation}
\chi^2=8\pi q^2 \lB c_0.
\end{equation}

The dielectric constant of the surface $\varepsilon'$ can be significantly different from the solvent $\varepsilon$. We define the dielectric contrast parameter as
\begin{equation}
\Delta=\frac{\varepsilon-\varepsilon'}{\varepsilon+\varepsilon'}.
\end{equation}
The presence of a large dielectric contrast at the interface ($\Delta\sim 1$, or $\varepsilon \gg \varepsilon'$) can lead to enhanced  effects due to the presence of polyvalent counterions~\cite{naji,jho-pre,kanduc-epje,jho-prl}. 

Note that no overcharging would be possible if the solution consisted only of neutralizing  counterions.
However, this situation is expected to be modified significantly by the presence of monovalent salt, where various degrees of overcharging of the charged surface have been reported previously~\cite{Lenz,Lozada,Molina, Kj, wang1, wang2}. 

\section{Dressed counterion theory}
In what follows we will apply the ``dressed counterion'' approach to our model system, which was introduced recently~\cite{kanduc-dressed} in a different context and was not thoroughly examined by explicit simulations. This approach is based on the fact that  for {\em highly asymmetric} solutions with $q\gg 1$, the monovalent ions can be  integrated out from the partition function and their effect is replaced by an effective (``dressed") DH interaction that acts between the remaining polyions and the surface charges. The idea stems from a realization that monovalent ions, because of their small valency, are only weakly coupled to all other charges, i.e., polyvalent counterions as well as surface charges. As a result, they can be integrated out, a procedure that can be justified systematically in a full field-theoretical framework~\cite{kanduc-dressed}. 

For the sake of simplicity, we shall illustrate this idea in the context of mean-field theory where the system is described via the Poisson-Boltzmann equation,
\begin{equation}
\nabla^2\psi=-4\pi\lB\left[n_0\rme^{-\psi}-(n_0 +qc_0)\rme^\psi+q c_0\rme^{-q\psi}+\tilde\rho_0(\Av r)\right],
\label{eq:PB}
\end{equation}
where the first three terms correspond to $+1$ ions, $-1$ ions and $+q$ ions, respectively. The fourth term $\tilde\rho_0(\Av r)$ stands for the external surface charges (in units of $e_0$) and $\psi$ is the dimensionless mean electrostatic potential (in units of $k_\textrm{B}T/e_0$). 
In the situation when $\psi\ll 1$ and $q \gg 1$, one can linearize the first two terms in the above PB equation as
\begin{equation}
\nabla^2\psi=\kappa^2\psi-4\pi\lB \left[qc_0\rme^{-q\psi}-qc_0+\tilde\rho_0(\Av r)\right].
\label{eq:PBlin}
\end{equation}
The constant zeroth-order term,  $-qc_0$, can be absorbed into the definition of the electrostatic potential $\psi$,  corresponding to a displacement of the potential by a constant Donnan potential.  This procedure changes nothing for our subsequent analysis.
We have hence introduced the screening parameter
\begin{equation}
\kappa^2=8\pi\lB(n_0+\tfrac 12 qc_0),
\label{eq:kappa}
\end{equation}
which is defined by the total amount of all monovalent ions, {\em i.e.}, ions coming from the 1:1 salt with concentration $n_0$ as well as the additional monovalent coions that enter the solution via the polyvalent $q$:1 salt with concentration $qc_0$. This form of the effective screening parameter follows also from more general statistical mechanical grounds in the context of colloidal interactions in the presence of low concentration of salts~\cite{Chan}.
It is also amended compared to the definition in our previous work in Ref.~\cite{kanduc-dressed}, where the regime of relatively high  monovalent salt concentration was of main interest.
 
Note however that the above PB equation cannot properly treat polyvalent ions due to high correlation effects. We now make a crucial step, namely we only treat monovalent ions in the linearized PB sense, whereas polyvalent ions along with surface charge are considered explicitly, expressed in terms of explicit charge density $\tilde\rho_\textrm{exp}(\Av r)$, viz.
\begin{equation}
\nabla^2\psi=\kappa^2\psi-4\pi\lB \tilde\rho_\textrm{exp}(\Av r).
\label{eq:DHmono}
\end{equation}
This linearized PB equation then leads straightforwardly to the relevant Green's function or the non-homogeneous DH kernel, representing the  screened DH interaction (modified by the presence of any dielectric interfaces) between all the remaining explicit charges after the simple salt degrees of freedom are integrated out. It can be represented for the present system {\em via} a 2D Fourier transform as
\begin{equation}
u_\textrm{DH}(Q;z,z')=\frac{1}{2\varepsilon\varepsilon_0 k}\left[\rme^{-k\vert z-z'\vert}+\frac{\varepsilon k-\varepsilon' Q}{\varepsilon k+\varepsilon' Q}\,\rme^{-k(z+z')}\right],
\label{eq:uDH}
\end{equation}
where $Q$ is the transverse wave-vector and $k=\sqrt{\kappa^2+Q^2}$. The first term is the usual (homogeneous) DH interaction in free space, and the second term is the non-homogeneous correction due to the presence of dielectric discontinuities and the inhomogeneous distribution of salt ions in the system (leading to screening discontinuities, i.e., no salt ions are present in the region $z<0$ $-$see Fig.~\ref{fig:geom}). 

Now in order to calculate the full partition of the system in any given setup (beyond the mere mean-field limit approximation that we used for the illustration above), one has to integrate over the degrees of freedom associated with the explicit polyvalent counterion charges as well (see Ref. \cite{kanduc-dressed} for a systematic field-theoretical treatment). We shall consider this procedure later for the study of the system in the strong-coupling limit. 

The dressed counterion approach thus enables an enormous reduction in the number of explicit particles in the system, comprised now only of polyvalent counterions and fixed surface charges that interact via ``dressed'' non-homogeneous DH interactions.  Although substantially simplified, the  ``dressed counterion" model still represents a full many-body problem (that can be used to study the system  in all regimes of the coupling parameter by prescribing the modified pair interaction according to Eqs.~(\ref{eq:uDH1}) and (\ref{eq:uDH1real})) and is as such exactly solvable only by computer simulations.

In many biologically relevant cases the dielectric material of the bounding surfaces has a dielectric constant $\varepsilon'\simeq 2-4$, as is the case with {\em e.g.} hydrocarbons or DNA hydrophobic core, whereas the solvent is water with $\varepsilon\simeq 80$. The notion that $\varepsilon\gg \varepsilon'$ ($\Delta\simeq1$) can simplify theoretical treatment enormously, since the non-homogeneous DH kernel, Eq.~(\ref{eq:uDH}), can in that case be rewritten as
\begin{equation}
u_\textrm{DH}(Q;z,z')=\frac{1}{2\varepsilon\varepsilon_0 k}\left[\rme^{-k\vert z-z'\vert}+\rme^{-k(z+z')}\right],
\label{eq:uDH1}
\end{equation}
or in real space simply as
\begin{equation}
u_{\rm DH}(\Av r,\Av r')=\frac{\rme^{-\kappa \vert\Av r-\Av r'\vert}}{4\pi\varepsilon\varepsilon_0\vert\Av r-\Av r'\vert}+\frac{\rme^{-\kappa \vert\Av r-\Av r'+2z'\Av e_z\vert}}{4\pi\varepsilon\varepsilon_0\vert\Av r-\Av r'+2z'\Av e_z\vert}.
\label{eq:uDH1real}
\end{equation}
This brings us to the concept of screened {\em image charges}, where the presence of each charge produces its own image on the other side of the interface (second term). Here $\Av e_z=(0,0,1)$ stands for unit vector perpendicular to the surface. In this limit the exact value of $\varepsilon'$ does not enter explicitly. If not stated otherwise we will mostly deal with the case where $\Delta\simeq 1$, which is also a very good approximation to the realistic $\Delta=0.95$ case as will be seen later, Fig.~\ref{fig:cSC}c. Note that the above form of the Green's function takes into account the standard dielectric images as well as the ionic cloud images stemming from the inhomogeneous distribution of salt in the system.

\subsection{Implicit vs. explicit simulations}

To examine the validity of the dressed counterion approach (for a highly asymmetric system with a large $q$) we first directly compare the results of implicit and explicit MC simulations. In order to perform the MC simulations with the dressed counterion {\em ansatz} that we also refer to as the {\em implicit MC simulations}, we need to evaluate the interactions between all the charged {\em via} the non-homogeneous DH kernel, Eqs.~(\ref{eq:uDH1}) and (\ref{eq:uDH1real}).

The interaction between every dressed counterion and the surface charge is then written as
\begin{eqnarray}
\beta W_{0c}&=&\beta \int\rho_0(\Av r)u_\textrm{DH}(\Av r,\Av r')\rho_c(\Av r')\rmd \Av r\rmd\Av r'\nonumber\\
			&=&-\frac{2}{\mu\kappa}\,\rme^{-\kappa z},
\label{eq:W0c}			
\end{eqnarray}
with surface charge operator $\rho_0(\Av r)=-\sigma\delta(z)$ and counterion charge operator $\rho_c(\Av r)=e_0q\delta(\Av R-\Av r)$ with $\Av R$ standing for the counterion position. 

The interaction between dressed counterions is additive. For every pair of different counterions $i$ and $j$, ($i\ne j)$, the interaction is
\begin{equation}
\beta W_{cc}(i,j)=\beta(e_0 q)^2\, u_\textrm{DH}(\Av r_i, \Av r_j).
\end{equation}
Each counterion itself also experiences a self-energy due to the dielectric half-space,
\begin{equation}
\beta W_\textrm{self}(z)=\frac{q^2 \lB}{4 z}\,\rme^{-2\kappa z}.
\label{eq:self}
\end{equation}
This repulsive interaction is responsible for the depletion effect in the vicinity of the surface. Its range is twice as short as attractive surface-charge interaction, Eq.~(\ref{eq:W0c}). The dielectric self-energy of the ion represents a generalization of the concept of  Born energy valid for an ion in a homogeneous dielectric. In the present paper we do not take into account the finite size of the ions in the dielectric self-energy, but we do take into account the presence of the dielectric inhomogeneities.

At this stage we introduce the dimensionless (rescaled) units by defining 
\begin{equation}
\Av{\tilde r}=r/\mu, \quad \tilde\kappa=\mu\kappa, \quad \tilde\chi=\mu\chi.
\end{equation}
In order to connect with unscaled values, table~\ref{table0} gives illustrative examples of actual concentrations for various dimensionless parameters $\tilde\kappa$ and $\tilde\chi$ for $\Xi=50$ and $\Xi=100$. 
Although higher values of dimensionless screening $\tilde\kappa>0.5$ can be difficult to achieve in aqueous solutions due to the high salinity required, it can be nevertheless more easily achieved in solvents with lower dielectric constant (e.g., ethanol).
\begin{table*}
	\begin{center}
\begin{tabular}{rrrrrrr|rrrrrr}
\hline\hline
&\multicolumn{6}{c|}{$\Xi=50$\rule{0pt}{3ex}}&\multicolumn{5}{c}{$\Xi=100$}\\
&$\tilde\kappa=0.1$\rule{0pt}{3ex}	&$0.2$	&	$0.5$	&	$0.8$&&&$\tilde\kappa=0.1$	&$0.2$	&	$0.5$	&	\\
\hline
$\tilde\chi=0.1$\rule{0pt}{3ex}	\vline&	$n_0=15$\,mM	&	\hspace{1ex}68\,mM	&	\hspace{1ex}440\,mM &	\hspace{1ex}1.1\,M 	&\vline&$c_0=1.1$\,mM&\hspace{1ex}$ n_0=63$\,mM		&	280\,mM	&	1.8\,M&\vline&$c_0=4.5$\,mM\\
$0.2$	\vline&	9\,mM	&	62\,mM	&	430\,mM &	1.1\,M	&\vline&4.7\,mM&36\,mM&250\,mM&1.8\,M&\vline&18\,mM\\
$0.5$	\vline&	$-$			&	15\,mM	&	390\,mM &	1.1\,M	&\vline&30\,mM&$-$&63\,mM&1.6\,M&\vline&110\,mM\\
$1.0$	\vline&	$-$			&	$-$		&	220\,mM &	0.9\,M	&\vline&120\,mM&$-$&$-$&0.9\,M&\vline&450\,mM\\
\hline\hline
\end{tabular}
\caption{Actual values of concentrations $n_0$ and $c_0$ in aqueous environment as follow from rescaled parameters $\tilde\kappa$ and $\tilde\chi$ for tetravalent counterions ($q=4$) and surface charge densities $\sigma=0.25\,e_0/$nm$^2$ ($\Xi=50$) and $\sigma=0.5\,e_0/$nm$^2$ ($\Xi=100$). Note that $c_0$ depends only on $\tilde\chi$, i.e., $c_0 = c_0(\tilde\chi)$, whereas $n_0 = n_0(\tilde\chi, \tilde\kappa)$ depends both on $\tilde\chi$ and $\tilde\kappa$.}
\label{table0}
\end{center}
\end{table*}

In our second approach we treat all the ionic species on the same footing by explicitly including them into MC simulations which we then refer to as the {\em explicit MC simulations}. For demonstration purposes we focus on the $q=4$ counterions and a surface charge density $\sigma=0.25\,e_0/$nm$^2$. In this case $\mu=0.23\,$nm whereas $\Xi=50$. In contrast, the coupling parameter for monovalent ions is $\Xi_\textrm{mono}=0.8$. To avoid Bjerrum pairing~\cite{vanroij} we have used closest-approach distances  $d_{+1,-1}=0.35$\,nm between monovalent cations and anions, and $d_{+4,-1}=0.45$\,nm between (polyvalent) counterions and monovalent coions.

\section{Results from implicit and explicit simulations}

\begin{figure*}[]
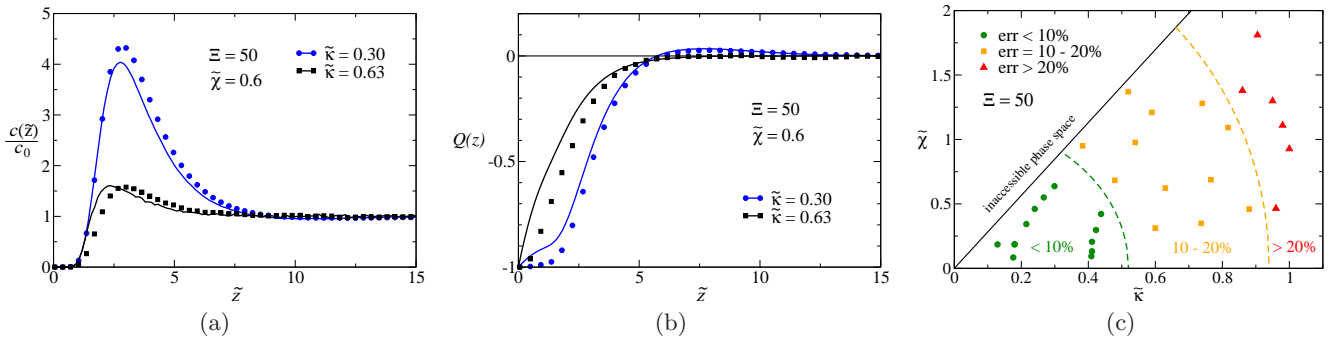
\begin{center}
	\begin{minipage}[b]{0.312\textwidth}\begin{center}
		\includegraphics[width=\textwidth]{cMCimpexp.eps} (a)
	\end{center}\end{minipage} \hskip0.25cm
	\begin{minipage}[b]{0.325\textwidth}\begin{center}
		\includegraphics[width=\textwidth]{overMCimpexp.eps} (b)
	\end{center}\end{minipage} \hskip0.25cm	
	\begin{minipage}[b]{0.302\textwidth}\begin{center}
		\includegraphics[width=\textwidth]{err.eps} (c)
	\end{center}\end{minipage} \hskip0.25cm	
\caption{(Color online) Comparison of explicit and implicit MC simulations for $\Xi=50$.
(a) Counterion density profile at fixed bulk amount of counterions $\tilde\chi$ and two different salt concentrations $\tilde\kappa$. Implicit MC results  (solid line) compare quite well with explicit MC results (dots).  (b) Cumulative charge that follows from the results in (a). In the case of smaller screening ($\tilde\kappa=0.30$) the cumulative charge reaches positive values, which corresponds to overcharging of $3\,\%$. (c)  ``Phase diagram" showing the deviations between implicit and explicit MC simulations for $\Xi=50$. The agreement is better for smaller values of $\tilde\kappa$.}
\label{fig:expimp}
\end{center}\end{figure*}

The results for (polyvalent) counterion density profiles obtained from both types of MC simulations are shown in Fig.~\ref{fig:expimp}a. Counterions are localized near the charged interface but are depleted from the immediate vicinity due to image repulsions. At large distances the profile saturates to the bulk value, $c_0$. The higher the monovalent salt concentration, i.e., the larger the screening $\tilde\kappa$, the smaller the amount of counterions close to the surface as larger amounts of salt screen the surface charge more efficiently. The comparison of explicit (dots) and implicit (solid line) MC simulations shows not only qualitative but also good quantitative agreement.

An interesting quantity to examine is the {\em cumulative charge}, $Q(z)$, being the sum of all charges within a finite distance from the surface, including the surface charges themselves 
\begin{eqnarray}
Q(z)&=&\frac {1}{\sigma} \int_0^z\left[e_0n_+(z')-e_0n_-(z')\right.\nonumber\\
&&\qquad +\left. e_0q c(z')-\rho_0(z')\right]\rmd z', 
\end{eqnarray}
where $n_\pm(z)$ represent the density of monovalent ions and $c(z)$ that of polyvalent counterions. 
Note that the $Q(z)$ is normalized so that it reaches  $Q =-1$ at the surface ($z=0$) and approaches $0$ far away from the surface ($z\to\infty$) due to complete neutralization of the surface charge.

In the case where the ions are treated  explicitly  the evaluation of $Q(z)$ is straightforward.
Within the implicit dressed counterion approach the monovalent salt densities in the definition of $Q$ need to be recalculated from the DH equation, viz. $n_+(\Av r)-n_-(\Av r)\simeq -2n_0\psi(\Av r)$, where $\psi$ corresponds to the dimensionless electrostatic potential that is evaluated from known explicit densities, {\em viz.}
\begin{equation}
\psi(\Av r)=\beta e_0\int u_\textrm{DH}(\Av r, \Av r')[e_0 q c(\Av r')+\rho_0(\Av r')]\, \rmd\Av r'.
\end{equation}
Inserting the surface charge density $\rho_0(\Av r)=-\sigma \delta(z)$ and the non-homogeneous DH kernel, Eq.~(\ref{eq:uDH1}), we get after some algebraic manipulation the expression for the cumulative charge in dimensionless units
\begin{eqnarray}
Q(z)&=&-\rme^{-\kappa z}+\frac 18 \tilde\chi^2\label{eq:Q}\\
&\times&\int_0^\infty\left[\textrm{sgn}(z-z')\rme^{-\kappa\vert z-z'\vert}+\rme^{-\kappa(z+z')}\right]\tilde c(\tilde z')\, \rmd \tilde z'.\nonumber
\end{eqnarray}
The first term is due to the DH exponential screening in the absence of polyions, whereas the second one corresponds to the effects due to polyvalent ions, where $\tilde c(z)=c(z)/c_0$.

 As seen in Fig.~\ref{fig:expimp}b the surface charge is neutralized more quickly in the case with a higher salt screening. If the screening is low enough, the surface can be overcharged leading to positive values of $Q$ at a finite $z$. The  overcharging degree turns out to be relatively small (in fact an order of magnitude smaller) compared to the magnitude of the bare surface charge, which agrees  quantitatively with results from several recent explicit MC studies on similar systems~\cite{wang1,wang2,Lenz} but disagrees with others~\cite{Nguyen1,Nguyen2,Nguyen3}. 
 
The agreement between implicit and explicit models is less pronounced close to the surface as the DH approach used in implicit simulations does not account for finite size of monovalent ions as well as image effects in  the distribution  of monovalent ions. Such discrepancies of ionic distributions in close vicinity to the interface however do not affect the cumulative charge at larger distances, where implicit and explicit results match perfectly. 

We shall quantify the discrepancy between implicit and explicit MC results by comparing the relative difference in the surface areas under the counterion profile curves. We thus calculate the integrals $\Delta S= \int |c_\textrm{imp}(z)-c_\textrm{exp}(z)|\rmd z$ and $S = \int c_\textrm{exp}(z)\rmd z$ within the interval corresponding to 2$\times$FWHM (full width at half maximum) on both sides of the peak. The relative error is then defined as $\textrm{err}=\Delta S/S$, and is presented in Fig.~\ref{fig:expimp}c. Note that according to the definition of $\kappa$ we should have $\tilde\chi\le \sqrt{2q}\,\tilde\kappa$. As can be seen in  Fig.~\ref{fig:expimp}c the agreement between implicit and explicit MC results is fairly good in a wide range of screening parameters. The agreement becomes less satisfying at large screenings and to some extent also when  $\tilde\chi$ is increased, {\em i.e.}, at larger concentrations of the polyvalent counterions. 

These results may seem in contradiction with a naive expectation that the DH approach performs better for larger screenings. Note however, that the finite size of monovalent ions becomes increasingly more important at 
higher screenings (salt concentrations), an effect not taken into account within the present DH approach. On the other hand, one might also expect a complete failure of the DH approach at low screenings, i.e., when $\tilde\kappa\lesssim 1/q$, especially near the charged surface where non-linear effects become important. In this case it turns out that the dielectric mismatch acts in the opposite direction and the image repulsion depletes the  counterions from the immediate vicinity of the surface, hence the exact form of the interaction with the surface charge in its vicinity is not important. Yet at very low screenings, some additional discrepancies might be observable due to non-linear effects. However, down to  $\tilde\kappa\simeq 0.1$ we still find very good agreement within $\textrm{err}<10\%$, c.f. Fig.~\ref{fig:expimp}c.

From Fig.~\ref{fig:expimp}c it thus appears that the dressed counterion theory is a good approximation for monovalent salt mediated interactions. This remains true in a very large segment of the parameter space where the discrepancies seldom exceed $20\%$, clearly corroborating the practical value of the dressed counterion approach (for highly asymmetric cases with $q\gg 1$) and the implicit simulations, which can be orders of magnitude less expensive than the standard explicit MC simulations, especially for very large systems of monovalent and polyvalent salt mixtures.

\section{SC dressed counterion limit}

We now focus on the analytical  SC description of our  system. As already stated, the monovalent and polyvalent ions need separate treatments due to their vastly different coupling to external charges. 
In the case of counterion-only systems the polyvalent counterions can be treated in the SC limit $\Xi\gg 1$ via a  virial expansion in terms of their fugacity $\lambda$ from the grand-canonical partition function ${\cal Z}_G$ as~\cite{Netz,hoda,naji} 
\begin{equation}
{\cal Z}_G={\cal Z}_G^{(0)}+\lambda{\cal Z}_G^{(1)}+\cdots.
\label{eq:ZG}
\end{equation}
Here the first term corresponds to the interaction of bare surface charges in the absence of any counterions, and the second term corresponds to the contribution of a single counterion in the system, etc. This approach turns out to be asymptotically exact for high values of the coupling parameter $\Xi$~\cite{Netz,hoda,naji}. In its original version the theory does not include any monovalent salt ions, being limited only to the effects of neutralizing counterions at charged surfaces~\cite{Netz}. Here we implement the same idea but take it a step further with ``dressed'' counterions to account for asymmetric systems containing additional salt (as in our previous work where two charged surfaces were studied  without any dielectric mismatches~\cite{kanduc-dressed}).
One can then follow a similar virial expansion approach as above and it turns out that the only change in the SC formulation wrought by the dressed counterions is to replace the standard Coulomb kernel in the original formulation~\cite{Netz,hoda,naji} by the non-homogeneous DH kernel, Eq.~(\ref{eq:uDH1}), in the dressed counterion formulation. This is then the essence of the {\em SC dressed counterion theory}. Note that in our case we focus on a grand-canonical system in equilibrium with a bulk solution (rather than a canonical system considered in the absence of salt screening~\cite{Netz,hoda,naji}), and hence the fugacity $\lambda$ is explicitly determined by the bulk concentration of (polyvalent) counterions $c_0$. 

From the first-order grand-canonical expansion, Eq.~(\ref{eq:ZG}), the grand potential $\beta\Phi=-\ln\,{\cal Z}_G$ of the dressed counterions can be evaluated as
\begin{equation}
\beta \Phi = -\ln\,{\cal Z}_G^{(0)}+\lambda\frac{{\cal Z}_G^{(1)}}{{\cal Z}_G^{(0)}}.
\end{equation}
As already mentioned the first-order term  ${\cal Z}_G^{(1)}$ stands for the one-particle dressed counterion contribution, hence
\begin{equation}
{\cal Z}_G^{(1)}={\cal Z}_G^{(0)}\int\exp\bigl(-\beta W_{0c}(\Av r)-\beta W_\textrm{self}(\Av r)\bigr)\, \rmd {\mathbf r}.
\label{eq:ZG1}
\end{equation}
The average number of dressed counterions is then obtained from the thermodynamic relation 
\begin{equation}
\overline N=-\lambda(\partial \beta\Phi/\partial\lambda).
\end{equation}
in complete analogy with the standard SC formulation for  ``bare" counterions. The dressed counterion density profile can be extracted from the form of the first-order term, Eq.~\ref{eq:ZG1}, and can be written explicitly
\begin{equation}
c(z) = c_0 \exp\left(\frac{2}{\>\tilde\kappa}\,\rme^{-\tilde\kappa \tilde z}-\frac{\Xi}{4\tilde z}\,\rme^{-2\tilde\kappa \tilde z}\right),
\label{eq:cSC}
\end{equation}
again analogous to the expression for the bare counterions. The first term on the r.h.s. corresponds to attractive surface-counterion interaction, Eq.~(\ref{eq:W0c}), and decays exponentially with the screening parameter $\tilde\kappa$. The second term is due to repulsive self-energies of counterions, Eq.~(\ref{eq:self}), having their origin in counterion self-image interactions. It decays twice as fast as the first one and becomes negligible at larger separations compared with the first term. However, at small separations it diverges as $1/\tilde z$ and hence engenders a depletion layer of counterions. Its strength increases with increasing coupling parameter $\Xi$ as is known from other image-interaction studies~\cite{kanduc-epje, jho-prl}.

At this stage we should emphasize that our SC dressed counterion limit, which by definition corresponds to the leading order in the virial expansion, does not necessarily correspond to a laterally correlated counterion cloud as is the case for the standard counterion-only systems. This fact should be taken into account when the validity of the SC limit will be considered, see below.

\begin{figure*}[]
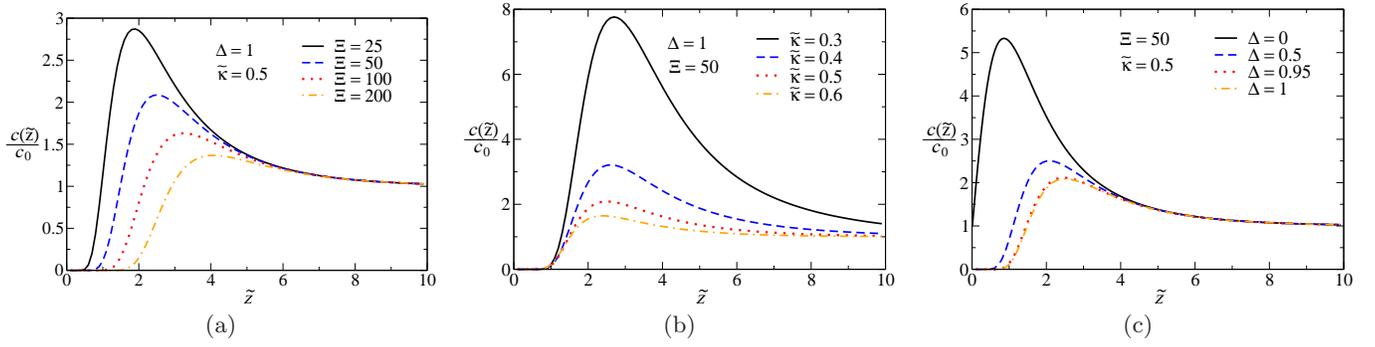
\begin{center}
	\begin{minipage}[b]{0.32\textwidth}\begin{center}
		\includegraphics[width=\textwidth]{cSC-Xi.eps} (a)
	\end{center}\end{minipage} \hskip0.25cm
	\begin{minipage}[b]{0.32\textwidth}\begin{center}
		\includegraphics[width=\textwidth]{cSC-k.eps} (b)
	\end{center}\end{minipage} \hskip0.25cm	
	\begin{minipage}[b]{0.32\textwidth}\begin{center}
		\includegraphics[width=\textwidth]{cSC-D.eps} (c)
	\end{center}\end{minipage} \hskip0.25cm	
\caption{(Color online) SC dressed counterion profiles for (a) various coupling parameters, (b) various screening parameters, and (c) various dielectric contrast parameters.}
\label{fig:cSC}
\end{center}\end{figure*}

In Fig.~\ref{fig:cSC} we show the main SC results and their trends in terms of all three adjustable parameters, $\Xi$, $\tilde\kappa$, and $\Delta$. As seen from Fig.~\ref{fig:cSC}a  increasing the coupling parameter $\Xi$ results in a large image repulsion, Eq.~(\ref{eq:self}), and hence a lower density at the peak. Similarly, the density peak decreases also by increasing the screening parameter $\tilde\kappa$ as shown in Fig.~\ref{fig:cSC}b. Thus the larger the screening, the weaker would be the interaction between all the explicit charges and therefore  a smaller amount of counterions is attracted towards the surface.

 Finally, we also show the density profiles for various values of the dielectric contrast $\Delta$, which requires an exact treatment of the non-homogeneous DH kernel, Eq.~(\ref{eq:uDH}) (rather than the simplified form (\ref{eq:uDH1})). 
The counterion-surface term $W_\textrm{0c}$, Eq.~(\ref{eq:W0c}), has the same form also when the exact kernel is taken into account, and is hence independent of the dielectric contrast. In contrast, the self-energy interaction $W_\textrm{self}$ takes a more complicated form for arbitrary $\Delta$, {\em viz.}
\begin{equation}
\beta W_\textrm{self}(\Av r)=\frac 12 \Xi\int_0^\infty\frac{\varepsilon \tilde k-\varepsilon'\tilde Q}{\varepsilon \tilde k+\varepsilon '\tilde Q}\,\frac{\tilde Q\rmd \tilde Q}{\tilde k}\,\rme^{-2\tilde k\tilde z},
\end{equation}
with $\tilde k=\sqrt{\tilde\kappa^2+\tilde Q^2}$. As can be seen from numerical results in Fig.~\ref{fig:cSC}c,  increasing the dielectric contrast results in  an increased  depletion and peak lowering, similar to the case where $\Xi$ or $\tilde\kappa$
are increased. Note also that in the case of a vanishing dielectric contrast, $\Delta=0$ or $\varepsilon'=\varepsilon$, there remains slight depletion of counterions from the surface vicinity, which is due to solvation effects of implicit salt. The latter situation corresponds to pure solvent on the left side (with $\kappa'=0$) and electrolyte solution on the right (with $\kappa>0$), which causes a free energy loss when every solvated counterion approaches the interface. It should be noted also that the results with the dielectric contrast $\Delta=0.95$ [corresponding $\varepsilon=80$ and $\varepsilon'=2$] almost match with the limiting case of $\Delta=1$, which justifies our previous approximations of the non-homogeneous DH kernel, Eq.~(\ref{eq:uDH1}).

\begin{figure*}[]
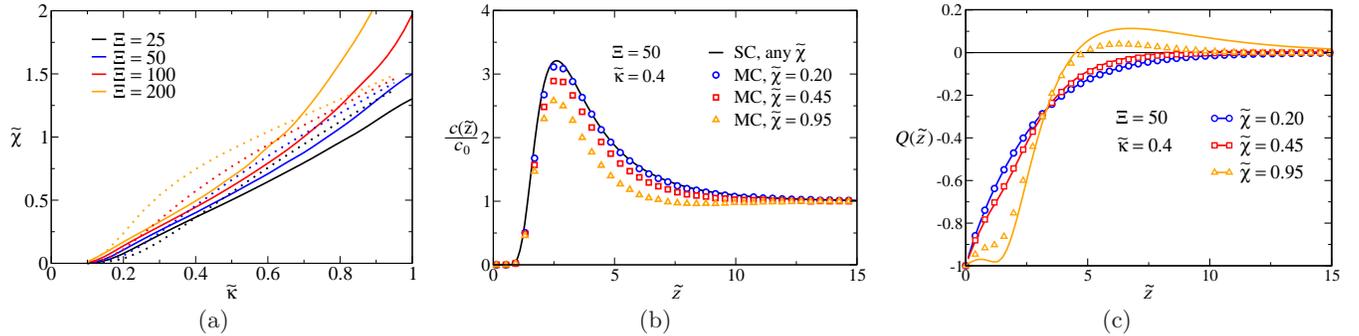
\begin{center}
	\begin{minipage}[b]{0.302\textwidth}\begin{center}
		\includegraphics[width=\textwidth]{OC-SC.eps} (a)
	\end{center}\end{minipage} \hskip0.25cm	
	\begin{minipage}[b]{0.313\textwidth}\begin{center}
		\includegraphics[width=\textwidth]{cSCimp.eps} (b)
	\end{center}\end{minipage} \hskip0.25cm
	\begin{minipage}[b]{0.332\textwidth}\begin{center}
		\includegraphics[width=\textwidth]{overSCimp.eps} (c)
	\end{center}\end{minipage} \hskip0.25cm	
\caption{(Color online) (a) ``Phase diagram" showing the minimal amount of polyvalent salt $\tilde\chi$ that is required to achieve overcharging (solid lines) within the SC dressed counterion theory. Dotted lines correspond to the situation where $\beta W_{cc}^\textrm{mf}=1$, Eq.~(\ref{eq:Wccmf}), with same colors correponding to the same parameters as shown on the graph. Parameter space below dotted lines should be well described by the SC dressed counterion approach. (b) Rescaled density profile for dressed counterions for $\Xi=50$ and screening $\tilde\kappa=0.4$. The SC prediction is independent of $\tilde\chi$ and gives good results for low enough $\tilde\chi$. With increasing $\tilde\chi$ the implicit MC results (symbols) show gradual deviations from the theoretical prediction. (c) Cumulative charge, Eq.~(\ref{eq:Q}), as follows from the profiles in (b) for the SC theory (solid lines) and implicit MC simulations (symbols).}
\label{fig:SCimp}
\end{center}\end{figure*}

It is instructive to look at the minimal value of the counterion concentration $c_0^\textrm{min}$ neccessary for overcharging. Fig.~\ref{fig:SCimp}a shows the minimal $\tilde\chi$ as a function of $\tilde\kappa$ (solid lines) which is necessary for overcharging. Above these lines the surface is overcharged whereas below it is not. Note the nearly linear dependence on $\Xi$ with a very mild slope  for $\tilde\kappa\lesssim 0.6$.

\subsection{SC criterion}

As already mentioned the SC dressed counterion approximation, Eq.~(\ref{eq:cSC}), differs from the standard counterion-only SC theory~\cite{Netz,naji,hoda} in two important aspects: the short-range nature of the dressed DH interaction and the grand-canonical ensemble used for  counterions. These differences engender several important modifications in the nature and interpretation of results. First of all it follows that on increase of the coupling parameter $\Xi$ one does not necessary approach the strong coupling limit as understood for counterion-only  systems. Instead, being a single particle theory with variable number of particles, it is expected to be valid when the interactions between counterions are negligible compared to thermal energy. This leads to a validity criterion which can be determined as follows. The effective interaction energy of a counterion at position $\Av r$ with all the other counterions can be estimated in a mean-field fashion as 
\begin{equation}
\beta W_{cc}^\textrm{mf}(\Av r)\simeq \beta(e_0q)^2\int u_\textrm{DH}(\Av r,\Av r')c(\Av r')\rmd \Av r'.
\label{eq:Wccmf}
\end{equation}
The highest contribution is expected for $\Av r$ around the profile peak, therefore in order to estimate the maximal $\beta W_{cc}^\textrm{mf}$ we choose $\Av r$ such that it correponds to the position of the maximum of $c(\Av r)$. The SC approach, Eq.~(\ref{eq:cSC}), is thus expected to be valid when $\beta W_{cc}^\textrm{mf}\ll 1$. Since this interaction grows with $c_0$, the SC approach is satisfied when $\tilde\chi$ is small enough. We plot   $\tilde\chi$ as a function of $\tilde\kappa$ that corresponds to $\beta W_{cc}^\textrm{mf}=1$ in Fig.~\ref{fig:SCimp}a (dotted lines). In the regions below these lines the SC theory is expected to work well. Here equal colors of solid and dotted lines correspond to the same $\Xi$. Note that solid and dotted lines roughly coincide. This brings us to the rule-of-thumb conclusion that the SC theory (\ref{eq:cSC}) is valid below the overcharging regime. With the emergence of overcharging the SC theory then starts to break down.

The above argument can be confirmed by comparing the SC prediction with MC simulations, Figs.~\ref{fig:SCimp}b,~\ref{fig:SCimp}c. Note that the rescaled density $c(z)/c_0$ is independent of $\tilde\chi$ in the SC theory, Eq.~(\ref{eq:cSC}).  As expected the comparison with the implicit MC data shows growing disagreement with the SC dressed counterion theory as $\tilde\chi$ is increased. The disagreement stems from the fact that the theory overestimates the counterion density as it neglects the repulsive counterion-counterion interactions and therefore predicts a larger number of counterions near the surface as compared with simulations.  However, the profile continues to display the same qualitative features, i.e., a peak with a depletion region, for a very wide range of $\tilde\chi$. The corresponding cumulative charge $Q(z)$, shown in Fig.~\ref{fig:SCimp}b, shows perfect agreement between the SC theory and simulations for the cases below the overcharging regime. The SC theory starts to break down quantitatively when the system approaches the overcharging point. The deviations in Fig.~\ref{fig:SCimp}c become significant for $\tilde\chi=0.95$.  This general conclusion can be deduced not only from the cumulative charge but also from the counterion density profiles and the "phase diagram", which confirms our rule of thumb that the SC prediction is valid below the overcharging regime, see below.

\begin{figure}[h]\begin{center}
	\begin{minipage}[b]{0.4\textwidth}\begin{center}
		\includegraphics[width=\textwidth]{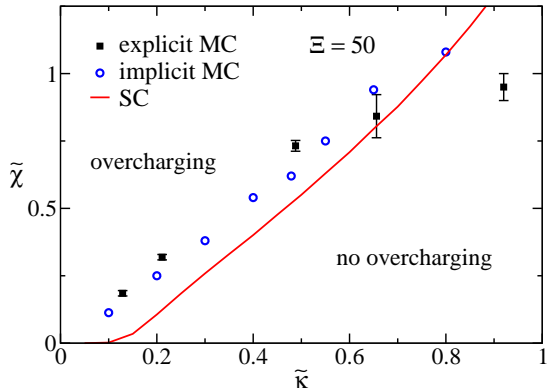} 
	\end{center}\end{minipage} 
\caption{(Color online) ``Phase diagram" characterizing the overcharging of a charged surface. The threshold values of minimal $\tilde\chi$ neccessary for overcharging are obtained by all three different methods, i.e., explicit and implicit MC simulations as well as the SC dressed counterion theory.}
\label{fig:OC}
\end{center}\end{figure}

Finally we compare the overcharging threshold value of $\tilde\chi$ from simulations and from SC dressed counterion theory. The  diagram in Fig.~\ref{fig:OC} shows the demarcation line between the overcharging regime and the absence thereof (for $\Xi=50$) as follows from all the three different approaches we have used in this study. The implicit and explicit MC simulations agree very well for screening parameters $\tilde\kappa<0.6$, which is in accordance with Fig.~\ref{fig:expimp}c. On the other hand, the results from the SC theory show a shift by a constant value towards a smaller threshold value for $\tilde\chi$. The reason is again that the SC theory overestimates the  counterion density  near the surface, and hence predicts a smaller $\tilde\chi$ needed for achieving the overcharging. Nevertheless, all three methods predict a clear and consistent trend: a nearly linear dependence of $\tilde\chi$ on $\tilde\kappa$ at the demarcation line. This fact was also observed experimentally, i.e., the more salt  is present in the system, the higher concentration of polyvalents is needed in order to observe overcharging~\cite{heyden}. As already seen in Fig.~\ref{fig:SCimp}a the SC theory gives a trend that is only weakly dependent on $\Xi$, which appears to indicate an approximate proportionality relation as $\tilde\chi^\textrm{min}\sim\tilde\kappa$. In real units this translates into a {\em heuristic relation} as
\begin{equation}
c_0^\textrm{min}\sim n_0/q^2,
\end{equation}
consistent with the experimentally observed fact that counterions with higher valency $q$ are more efficient at overcharging, which has also been observed experimentally~\cite{besteman}. Nevertheless one needs to keep in mind that ion-specific effects~\cite{benyaakov} also make their mark in strongly coupled systems. The SC theory based on purely electrostatic grounds appears to be unable to account for the experimentally observed variation in the strength of the SC effects for equally charged polyvalent counterions~\cite{todd}. The ion specific effects in the SC theory have remained largely unexplored. 

\section{Conclusion}

In our work we present detailed arguments based on extensive simulation for the validity of the dressed counterion approach and its SC limit, based on a model for an ionic solution mixture composed of a monovalent (1:1) salt and an asymmetric ($q$:1) salt in proximity to a charged dielectric interface.

In the first part we have described the general idea of dressed counterions, i.e., the coarse-grained electrostatic treatment that replaces explicit monovalent-ion degrees of freedom by the effective Debye-H\"uckel interaction between the remaining charges comprised of the polyvalent counterions and the surface charges. This idea is valid in highly asymmetric systems with $q\gg 1$ and stems from the fact that monovalent ions are weakly coupled with all the other charges. We compared the MC results of dressed counterions (referred to as implicit simulations) with the standard explicit MC simulations (referred to as explicit simulations) where all the ions are treated explicitly. Comparing the results for the density profiles of polyvalent counterions we found excellent agreement between both types of simulations.
 

Generally, counterions are attracted toward the surface because of direct electrostatic interactions but are expelled from the vicinity of the surface due to the dielectric image charges and in general solvation repulsion. For large polyvalent salt concentration we observed overcharging, i.e., the net excess of opposite charge near the surface. Both types of simulations agree fairly well on the degree of overcharging, which is typically of the order of few percent of the surface charge, in accordance with recent MC studies~\cite{wang1,wang2,Lenz}.

In the second part of the paper we implemented the analytical SC dressed counterion approach, based on a strong-coupling one-particle description~\cite{Netz}, to the model at hand. The SC dressed counterion theory gives a very simple analytical expression for the counterion density profile, Eq.~(\ref{eq:cSC}), and also enables  simple predictions for the onset of overcharging. We estimated the regime of validity of the SC theory based on the fact that the counterion-counterion interactions should be negligible when compared to the thermal energy. As it turns out this criterion approximately coincides with the onset  of overcharging. In other words,  the SC dressed counterion theory is valid in the regime just below the overcharging regime, which was confirmed by implicit MC simulations. In its regime of validity the SC dressed counterion theory gives good  predictions for the density profile of counterions and can also approximately predict the onset of overcharging. In agreement with the MC simulations, the theory shows a linear relationsip between the polyvalent salt concentration needed for overcharging and the monovalent salt concentration in the system. It also predicts higher-valency counterions are more effective in overcharging the surface. These results agree with recent  experimental findings~\cite{heyden,besteman}.

Finally we should emphasize that the merit of the dressed counterion approach is firstly in its practical use in implicit MC simulations, which give satisfactory results and can be several orders of magnitude less expensive compared to explicit MC simulations. Secondly, it can be easily included in analytical frameworks for charged systems (e.g., the strong coupling theory in our case), which may thus allow for a deeper understanding of the general features of the charged systems that contain highly asymmetric electrolytes.

\section{Acknowledgement}
R.P. acknowledges support from the Agency for Research and Development of Slovenia (Grants P1-0055(C)), M.K. is supported by the Alexander von Humboldt Foundation. A.N. is supported by the Royal Society, the Royal Academy of Engineering, and the British Academy. J.F. acknowledges support from the Swedish Research Council. This work was completed at Aspen Center for Physics during the workshop on {\em New Perspectives in Strongly Correlated Electrostatics in Soft Matter} (2010), organized by Gerard C.L. Wong and Erik Luijten.

\end{document}